\documentstyle[aps,prd,12pt,tighten,amssymb,amsmath,eqsecnum,cite]{revtex}

\def\bea{\begin{eqnarray}}
\def\eea{\end{eqnarray}}
\def\ga{\gamma}
\def\ep{\epsilon}

\title{{\large{\bf
Legendre's Relation and the Quantum Equivalence \\
osp$(4|4)_1$ $=$ osp$(2|2)_{-2}$ $\oplus$ su$(2)_0$}}}
\vspace{0.5cm}
\author{Miraculous J. Bhaseen\footnote{\vspace*{-0.4cm}bhaseen@thphys.ox.ac.uk}
\\ \vspace{0.5cm}
{\small {\em Physics Department, University of Oxford, Theoretical Physics, \\ 1 Keble Road,
Oxford, OX1 3NP, U.K.}}}
\begin{document}
\maketitle
%% To add preprint numbers:  galley style
%!l
%\vspace*{-9.5cm}
%\begin{flushright}
%  OUTP-99-07S\\
% cond-mat/9912060
%\end{flushright}
%\vspace*{8cm}

\begin{center}
\today
\end{center}
\vspace{0.5cm}

\begin{abstract}
Using explicit results for the four-point correlation functions of the
Wess--Zumino--Novikov--Witten (WZNW) model we discuss the conformal embedding osp$(4|4)_1$ $=$ osp$(2|2)_{-2}$ $\oplus$
su$(2)_0$. This embedding has emerged in Bernard and LeClair's recent
paper \cite{Bernard:spincharge}. Given that the osp$(4|4)_1$ WZNW model is a free theory with power law
correlation functions, whereas the su$(2)_0$ and
osp$(2|2)_{-2}$ models are CFTs with logarithmic correlation
functions, one immediately wonders whether or not it is possible to combine these logarithms 
and obtain simple power laws. Indeed, this very concern has been raised
in a  revised version of \cite{Bernard:spincharge}. In this paper we
demonstrate how one may recover the free field behaviour from a
braiding of the solutions of the  su$(2)_0$ and osp$(2|2)_{-2}$
Knizhnik--Zamolodchikov equations. We do this by implementing a procedure analogous to
the conformal bootstrap programme \cite{BPZ}. Our ability to recover such simple
behaviour relies on a remarkable identity in the theory of elliptic integrals known as Legendre's relation.  
\end{abstract}

%PACS: 72.15.Rh, 71.30.+h

%Key words: Spin Quantum Hall Effect, Conformal Symmetry.

%No preprint numbers for preprint style.

\section{Introduction}
Since the pioneering work of Belavin, Polyakov and Zamolodchikov \cite{BPZ}, two-dimensional conformally
invariant field theories have attracted a great deal of
attention (see for example \cite{Francesco:CFT}). A prominent r{\^ o}le in this arena is played by the WZNW
models \cite{WessZumino,Novikov,Witten:nonab} posessing additional Lie algebraic symmetry.  In
condensed matter physics, these models have been used to
describe the critical points of many low-dimensional physical systems.
In particular, in the theory of disordered systems, averaging over disorder may lead
to field theories defined over supermanifolds \cite{Efetov:chaos} and it
is natural to expect that WZNW models based on Lie superalgebras\footnote{For more information on Lie superalgebras we
refer the reader to the work of Kac \cite{Kac:class} and the
dictionary on superalgebras \cite{Frappat:dict}.} may describe their critical points
\cite{Bernard:perturbed,Maassarani:non,Bernard:spincharge}. In
addition, averaging over disorder, may lead to the so-called logarithmic
conformal field theories (LCFTs)
\cite{Gurarie:Log,Gurarie:Conf,Cardy:Logs}. A recent example of
such supersymmetric LCFTs has been considered in a paper by Bernard and LeClair \cite{Bernard:spincharge}. An essential
ingredient in their analysis is the observation that the Sugawara
energy-momentum tensor for the {\rm osp}$(4|4)_1$ Lie superalgebra,
separates into two commuting pieces:
\begin{equation}
\label{sugsep}
T_{\rm{osp}(4|4)_1}=T_{\rm{osp}(2|2)_{-2}}+T_{{\rm su}(2)_0}
\end{equation}
The conformal field theories appearing on the left and right hand
sides of this expression are quantum
equivalent, if and only if, they posess the same operator
algebra.\footnote{For a discussion of
quantum equivalence see \S 3.7 of the book by Fuchs
\cite{Fuchs:affine}.} As the authors of \cite{Bernard:spincharge} are acutely aware, the
decomposition (\ref{sugsep}) is a necessary, {\em but not sufficient},
condition to ensure this quantum equivalence. In addition to (\ref{sugsep})
one also requires the coincidence of the four-point
functions of primary fields \cite{Fuchs:affine}. However, the 
osp$(2|2)_{-2}$ and su$(2)_0$
 WZNW models are known to contain logarithmic
operators 
\cite{Caux:suzero,Kogan:origin,Maassarani:non} whereas the
osp$(4|4)_1$ WZNW model is known have power law correlation
functions \cite{Bernard:spincharge}. One immediately enquires whether
or not it is possible to combine these logarithms and obtain simple
power laws. In this paper we demonstrate that it is in fact possible
to combine these logarithms and achieve a reversion to
power law behaviour.

The structure of this paper is as follows: in section \ref{su2} we
study the four-point functions of the su$(2)_0$ WZNW
model.
 We note that these conformal blocks
may be expressed in terms of the complete elliptic integrals --- a
representation which turns out to be extremely fruitful. In section
\ref{osp} we study the four-point correlation functions of the
so-called $[0,1/2]$ representation of osp$(2|2)$
\cite{Scheurnert:ospm,Scheurnert:ospm2}. The Knizhnik--Zamolodchikov
equation for this representation has been provided by Maassarani and
Serban \cite{Maassarani:non}. As we discuss more fully in
\cite{Bhaseen:osp22}, the conformal blocks undergo a dramatic simplification at $k=-2$, and it is possible to implement the
conformal bootstrap \cite{BPZ} on a reduced set of solutions to the
Knizhnik--Zamolodchikov equation. Remarkably, this reduced subset
assumes a simple form in terms of the complete elliptic integrals. In section \ref{Legendre} we
demonstrate how Legendre's relation \cite{Legendre:Calcul,Whittaker:Modern} enables one to braid the conformal blocks of sections
\ref{su2} and \ref{osp} so as to reproduce the free field behaviour of
osp$(4|4)_1$. Finally we present concluding remarks and technical appendices.
\section{The su$(2)_0$ WZNW model}
\label{su2}
Following Knizhnik and Zamolodchikov \cite{Knizam:current} we consider
the four-point functions of the WZNW model
\begin{equation}
{\mathcal F}^{\alpha,{\bar{\alpha}}}(z_i,{\bar z}_i) = \langle g^{\alpha_1,{\bar \alpha}_1}(z_1,{\bar
z}_1)g^{\dagger \alpha_2,{\bar \alpha}_2}(z_2,{\bar
z}_2)g^{\dagger \alpha_3,{\bar \alpha}_3}(z_3, {\bar z}_3)g^{\alpha_4,{\bar \alpha}_4}(z_4,{\bar z}_4)\rangle
\end{equation}
where $g$ transforms in the fundamental representation of su$(N)$ and
$g^{\dagger}=g^{-1}$ transforms in the conjugate
representation. We use the symbol $\alpha$ to denote the ordered sequence
of su$(N)$ tensor indices $\alpha_1,\alpha_2,\alpha_3,\alpha_4$. Conformal invariance allows us to write
\begin{equation}
\label{su2corr}
{\mathcal F}^{\alpha, \bar \alpha}(z_i,{\bar z}_i) = [z_{14}z_{23}{\bar z}_{14}{\bar
z}_{23}]^{-2h}F^{\alpha, \bar \alpha}(z,\bar z)
\end{equation}
where $h$ is the conformal dimension the field $g(z,\bar z)$ (which in
the case of su$(2)_0$ is $1/8$), $z_{ij}=z_i-z_{j}$,
%\begin{equation}
%h={\bar h} = \frac{N^2-1}{2N(k+N)}
%\end{equation}
and the anharmonic ratios $z$ and $\bar z$ are defined as
\begin{equation}
\label{anhar}
z=\frac{z_{12}z_{34}}{z_{14}z_{32}}, \quad {\bar z}=\frac{{\bar
z}_{12}{\bar z}_{34}}{{\bar z}_{14}{\bar z}_{32}}
\end{equation}
The correlation function (\ref{su2corr}) admits the
invariant decomposition
\begin{equation}
F^{\alpha, \bar \alpha}(z,\bar z)=\sum_{ij=1}^mI_i^\alpha{\bar I}_j^{\bar
\alpha}F_{ij}(z,\bar z)
\end{equation}
with the $m=2$ su$(N)$ invariant tensors $I_1=\delta_{\alpha_1,\alpha_2}\delta_{\alpha_3,\alpha_4}$ and $I_2=\delta_{\alpha_1,\alpha_3}\delta_{\alpha_2,\alpha_4}$.
The four scalar functions $F_{ij}$ satisfy the coupled first order
equations
\begin{equation}
\label{su2matrixde}
\frac{dF}{dz}=\left[\frac{1}{z}P+\frac{1}{z-1}Q\right]F 
\end{equation}
where $F$ denotes the $2\times 2$ matrix $F_{ij}$, and the matrices
$P$ and $Q$ are given by
\begin{equation}
\label{su2pq}
P=\frac{1}{2N\kappa}\begin{pmatrix} N^2 -1 & N \\ 0 & -1
\end{pmatrix}, \quad Q = \frac{1}{2N\kappa}\begin{pmatrix} -1 & 0 \\ N & N^2-1
\end{pmatrix}
\end{equation}
where $\kappa=-\frac{1}{2}(N+k)$. In order to study the case su$(2)_0$ we set $N=2$ and $k=0$ in
equation (\ref{su2pq}). Supressing the
antiholomorphic index $j$ from the entries of $F$, one may reduce the
first-order differential equation (\ref{su2matrixde}) in this case to the following
pair of equations
\begin{subequations} 
\begin{gather}
\label{su2f1}
[4z(1-z)]^2F_1^{''}+(3-4z)\left[8z(1-z)F_1^{'}-F_1\right]=0 \\
\label{su2f2}
F_2=-2zF_{1}^{'}-\frac{(3-2z)}{2(1-z)}F_1
\end{gather}
\end{subequations}
It is easily seen that (\ref{su2f1}) admits the two solutions\footnote{Upon the change of variables
$F_1(z)=[z(1-z)]^{1/4}H(z)$
equation (\ref{su2f1}) reduces to the canonical form of the
hypergeometric equation $z(1-z)H^{''}+[c-(a+b+1)z]H-abH=0$
with $a=1/2$, $b=3/2$ and $c=2$. This has solutions
$H^{(1)}= \ _{2}F_{1}[\tfrac{1}{2}, \tfrac{3}{2};2;z]$ and 
$H^{(2)}= \ _{2}F_{1}[\tfrac{1}{2}, \tfrac{3}{2};1;1-z]$ which may be
expressed in terms of the elliptic integrals using the identities (\ref{su0soln1}) and (\ref{su0soln2}).}
\begin{subequations}
\label{su2ellipticsolutions}
\begin{eqnarray}
F_1^{(1)} &  = &  z^{-3/4}(1-z)^{1/4}\,[E(z)-K(z)] \label{su2f11}\\
F_1^{(2)}  & = &  z^{-3/4}(1-z)^{1/4}\,E(1-z)\label{su2f12}
\end{eqnarray}
\end{subequations}
where $K(z)$ is the complete elliptic integral of the first
kind and $E(z)$ is the complete elliptic integral of the second
kind as defined in equations (\ref{elliptichyper}) and (\ref{ellipticintegraldefs}).
%:\footnote{Note that many texts on the theory of elliptic integrals
%denote the parameter $z$ by $k^2$ - the so-called modulus. This is
%purely a mater of convention.}
%\begin{equation}
%E(z)=\int_0^1\frac{\sqrt{1-zx^2}}{\sqrt{1-x^2}}dx\hspace{1cm} K(z)=\int_0^1\f%rac{1}{\sqrt{(1-x^2)(1-zx^2)}}\, dx
%\end{equation}
The representation (\ref{su2ellipticsolutions}) is particularly useful\footnote{This
becomes even more apparant when we study the case of osp$(2|2)_{-2}$.} owing to the very
simple manner in which the elliptic integrals behave under
differentiation  with respect to the parameter $z$ --- see equation (\ref{simplediff}).
%\begin{equation}
%\label{simplediff}
%\frac{dE(z)}{dz}=\frac{E(z)-K(z)}{2z} \hspace{1cm} \frac{dK(z)}{dz}=\frac{E(z%)-(1-z)K(z)}{2z(1-z)}
%\end{equation}
Applying (\ref{su2f2}) to these solutions yields:
\begin{subequations}
\label{derivedsu2}
\begin{eqnarray}
F_2^{(1)} & = &  z^{1/4}(1-z)^{-3/4}\,E(z) \label{su2f21}\\
F_2^{(2)} & = &  z^{1/4}(1-z)^{-3/4}\,[E(1-z)-K(1-z)]\label{su2f22}
\end{eqnarray}
\end{subequations}
The results (\ref{su2ellipticsolutions}) and (\ref{derivedsu2}) agree with those of \cite{Caux:suzero,Kogan:origin}, but are presented in a
form which, for our purposes, is more convenient. In section \ref{osp}
we perform a similar analysis for the osp$(2|2)_{-2}$ WZNW model.  
\section{The osp$(2|2)_{-2}$ WZNW model}
\label{osp}
Following Maassarani and Serban \cite{Maassarani:non} we consider the
four-point functions of the WZNW model
\begin{equation}
\label{osp22corr}
{\mathcal F}^{\alpha,{\bar{\alpha}}}(z_i,{\bar z}_i) = \langle g^{\alpha_1,{\bar \alpha}_1}(z_1,{\bar
z}_1)g^{ \alpha_2,{\bar \alpha}_2}(z_2,{\bar
z}_2)g^{\alpha_3,{\bar \alpha}_3}(z_3, {\bar z}_3)g^{\alpha_4,{\bar \alpha}_4}(z_4,{\bar z}_4)\rangle
\end{equation}
where $g$ transforms in the so-called $[0,1/2]$ representation of the
Lie superalgebra osp$(2|2)$ and once again we use the symbol $\alpha$ to denote the ordered sequence
of indices $\alpha_1,\alpha_2,\alpha_3,\alpha_4$. The representation
theory of osp$(2|2)$
has been considered in \cite{Scheurnert:ospm, Scheurnert:ospm2}. In particular, the $[0,1/2]$
representation is four-dimensional and the index $\alpha_i$ takes on
the values $1,2,3,4$. In the notation of \cite{Maassarani:non}, the
indices $1,4$ are even (bosonic) while  $2,3$ are odd (fermionic).\footnote{For
reader unfamiliar with the Lie superalgebras we note that the Hilbert
space basis states (upon which matrix representations may be formed)
carry a so-called grading --- the state vectors themselves may be either
bosonic or fermionic.} As in section \ref{su2}, conformal invariance
allows us to write (\ref{osp22corr}) in the form (\ref{su2corr}) where this time
%\begin{equation}
%{\mathcal F}^{\alpha, \bar \alpha}(z_i,{\bar z}_i) = [z_{14}z_{23}{\bar z}_{1%4}{\bar
%z}_{23}]^{-2h}F^{\alpha, \bar \alpha}(z,\bar z)
%\end{equation}
$h=1/8$ for the field $g(z,\bar z)$
transforming under $[0,1/2]$ \cite{Maassarani:non}.
The correlation function (\ref{osp22corr}) also admits the invariant decomposition
%\begin{equation}
%F^{\alpha, \bar \alpha}(z,\bar z)=\sum_{ij=1}^3I_i^\alpha{\bar I}_j^{\bar
%\alpha}F_{ij}(z,\bar z)
%\end{equation}
where the $m=3$ osp$(2|2)$ invariant tensors derived in \cite{Maassarani:non} are reproduced in appendix \ref{invariant}. 
The nine scalar functions $F_{ij}$ satisfy the coupled first order
equations (\ref{su2matrixde}) where this time F denotes the $3\times 3$ matrix $F_{ij}$
and the matrices $P$ and $Q$ are given by
\begin{equation}
P=\frac{1}{x}\left( \begin{array}{ccc}
1 & 0 & 0 \\
-2 & -3 & -\frac{1}{2\ep\ga} \\
4\ep\ga & 8\ep\ga & 1    \end{array} \right), \quad
Q=\frac{1}{x}\left( \begin{array}{ccc}
-1 & 0 & -\frac{1}{2\ep\ga} \\
2 & 1 & \frac{1}{2\ep\ga} \\
-4\ep\ga & -4\ep\ga & -1 \end{array} \right) 
\end{equation}
where $x=2-k$. The two free parameters $\epsilon$ and $\gamma$ correspond to the
arbitrary relative normalizations of the su$(2)$ doublet ($|1\rangle$,
$|4\rangle$) and the two singlets $|2\rangle$ and $|3\rangle$ which
together provide a four-dimensional basis for the $[0,1/2]$
representation of osp$(2|2)$
\cite{Scheurnert:ospm,Scheurnert:ospm2}. Supressing the
antiholomorphic index  $j$ from the entries of $F$, one may reduce the
first-order differential equation (\ref{su2matrixde}) in this case to the following
set of equations \cite{Maassarani:non}
\begin{align}
\begin{split}
& x^3z^3(1-z)^3F_3^{'''}+x^2(1+2x)z^2(1-z)^2(1-2z)F_3^{''}+ \\ 
& \hspace{3cm} xz(1-z)[-1-x+2xz-2x(2+x)z(1-z)]F_3^{'} + \\
& \hspace{7cm} [-1-x+2z+2xz(1-z)]F_3(z)=0
\end{split} \label{f3} \\
\begin{split}
F_2(z) = -\frac{1}{4\epsilon\gamma
xz(1-z)}\left[x^2D^2F_3(z)+2x(1-z)DF_3(z)+(1-2z)F_3(z)\right] 
\end{split}\label{f2} \\
\begin{split}
F_1(z)=\frac{1}{4\epsilon\gamma}\left[xDF_3(z)-F_3(z)\right]+(z-2)F_2(z)
\end{split}\label{f1}
\end{align}
where $D=z(1-z)d/dz$. The authors of \cite{Maassarani:non} tackle (\ref{f3}) by introducing ancilliary
functions $F^{-}$ from which
one may obtain the $F_3$ via the differential relation $F_3\propto (1-xD)F^-$.
The $F^-$ satisfy a third-order equation, which
the authors of \cite{Maassarani:non}  solve in terms of generalized
hypergeometric functions --- see
equations 68-70 of
\cite{Maassarani:non}. One of these solutions is of the form\footnote{We note the small but significant typing error
in the last indicial argument of this function as it appears in
equation 68 of \cite{Maassarani:non}. The other two solutions 69 and
70 are correct, however.}
\begin{equation}
\label{fom}
F_0^{-}(z)= [z(1-z)]^{-1/x} \
_{3}F_{2}\left[\tfrac{1}{2},-\tfrac{1}{x},1-\tfrac{1}{x};1,1-\tfrac{2}{x};4z(1-z)\right].
\end{equation}
In particular, for the study of osp$(2|2)_{-2}$ we must set $x=4$. The {\em generalized} hypergeometric
function appearing in (\ref{fom}) reduces
under these circumstances to an {\em ordinary} hypergeometric
function. Using the well known identities satisfied by the ordinary
hypergeometric functions, one is able to obtain from this result, {\em
two}
%\footnote{Two quadratic transformations exist: $\
%_{2}F_{1}[\alpha,\beta;\alpha+\beta+\tfrac{1}{2};4z(1-z)]=\
%_{2}F_{1}[\alpha,\beta;\alpha+\beta+\tfrac{1}{2};x]$ where $x=z$ in
%the left lobe of the lemniscate $|4z(1-z)|=1$  and $x=1-z$ in the
%right --- see for example 4.\,(iii) and 4.\,(v) on page 97 of
%\cite{Bailey:ge%neralized}.} 
independent solutions to the equation (\ref{f3}) in terms of the
elliptic integrals (see Appendix \ref{quadraticosp}):
 \begin{equation}
\label{ellipticsolutions}
F_{3}^{(1)} = \frac{E(z)}{[z(1-z)]^{1/4}} \hspace{1cm} F_{3}^{(2)} = \frac{E(1-z)-K(1-z)}{[z(1-z)]^{1/4}}
\end{equation}
These solutions may be verified 
straightforwardly using (\ref{simplediff}). Applying (\ref{f2}) and (\ref{f1}) to these solutions yields:
\begin{alignat}{2}
F_{2}^{(1)} & =  -\frac{K(z)}{4\epsilon\gamma [z(1-z)]^{1/4}} 
& \qquad 
F_{1}^{(1)} & =  \frac{z K(z)}{4\epsilon \gamma [z(1-z)]^{1/4}} \label{blocks1}
\\
F_{2}^{(2)} & = \frac{K(1-z)}{4\epsilon \gamma [z(1-z)]^{1/4}} 
& \qquad 
F_{1}^{(2)} & = -\frac{z K(1-z)}{4 \epsilon \gamma [z(1-z)]^{1/4}}\label{blocks2}
\end{alignat}
 As we demonstrate in \cite{Bhaseen:osp22}, one may satisfy the demands of
single-valuedness and crossing symmetry --- the so-called conformal bootstrap
\cite{BPZ} --- on the subspace of functions
(\ref{ellipticsolutions})--(\ref{blocks2}). In section \ref{Legendre}
we shall demonstrate how one may braid the su$(2)_0$ conformal blocks
given by (\ref{su2ellipticsolutions}) and (\ref{derivedsu2}) with
those of the osp$(2|2)_{-2}$ model appearing in
(\ref{ellipticsolutions})--(\ref{blocks2}), so as to produce simple power laws.
\section{Legendre's Relation and osp$(4|4)_1$}
\label{Legendre}
\subsection{Embedding Blocks}
We begin by defining embedding blocks
\begin{equation}
E_{ij}(z)=\sum_{a,b=1}^2c_{ij}^{ab}\,F_{i,\,{\rm su}}^{(a)}(z)F_{j,\,{\rm osp}}^{(b)}(z)
\end{equation}
which are a simple braiding of the (holomorphic) su$(2)_0$ conformal blocks, $F_{i,\,{\rm su}}^{(a)}(z)$, and the (holomorphic) osp$(2|2)_{-2}$ conformal
 blocks, $F_{i,\,{\rm osp}}^{(a)}(z)$. The $c_{ij}^{ab}$ are, as yet,
 undetermined coefficients which we shall determine by imposing
 single-valuedness and crossing symmetry. 
In order to impose the condition of single-valuedness on the
embedding blocks, we consider the monodromy transformations of its
 constituent conformal blocks. A monodromy transformation of a function of z consists in letting z
circulate around some other point (typically a singular
point). We define
\begin{eqnarray}
{\mathcal C}_0\,F(z,\bar z) & = & \lim_{t\rightarrow 1^-}F(ze^{2i\pi
t},\bar ze^{-2i\pi t}) \\
{\mathcal C}_1\,F(z,\bar z) & = & \lim_{t\rightarrow 1^-}F(1+(z-1)e^{2i\pi t},1+(\bar z-1)e^{-2i\pi t})
\end{eqnarray}
Using the analytic continuation formulae for the hypergeometric
functions it is straightforward to see that
\begin{eqnarray}
{\mathcal C}_0\, F_{i,\,{\mathcal A}}^{(a)}(z) & = &
(g_0^{\mathcal A})_{ab}\,F_{i,\,{\mathcal A}}^{(b)}(z) , \quad i=1,2,3\\
{\mathcal C}_1\, F_{i,\,{\mathcal A}}^{(a)}(z) & = &
(g_1^{\mathcal A})_{ab}\,F_{i,\,{\mathcal A}}^{(b)}(z) , \quad i=1,2,3
\end{eqnarray}
where the algebra index ${\mathcal A}$ is either su or osp, and the
monodromy matrices $g_0^{\mathcal A}$ and $g_1^{\mathcal A}$ are given by
\begin{equation}
g_0^{\rm su}=\begin{pmatrix} i & 0  \\ -2 &
    i 
	\end{pmatrix}, \quad g_1^{\rm su}=\begin{pmatrix} i & -2  \\ 0 &
    i  \\ 
	\end{pmatrix}, \quad g_0^{\rm osp}=\begin{pmatrix} -i & 0  \\ 2 &
    -i 
	\end{pmatrix}, \quad g_1^{\rm osp}=\begin{pmatrix} -i & 2  \\ 0 &
    -i  \\ 
	\end{pmatrix}.
\end{equation}
We notice that these matrices are related by $g_i^{\rm su}=-g_i^{\rm
osp}$. Demanding that the embedding functions be invariant under both
${\mathcal C}_0$ and ${\mathcal C}_1$, requires that
$c_{ij}^{11}=c_{ij}^{22}=0$ and $c_{ij}^{12}=-c_{ij}^{21}$ ($=c_{ij}$
say.) In other words,
\begin{equation}
\label{eij}
E_{ij}(z)=c_{ij}\left[F_{i,\,{\rm su}}^{(1)}(z)F_{j,\,{\rm osp}}^{(2)}(z)-F_{i,\,{\rm su}}^{(2)}(z)F_{j,\,{\rm osp}}^{(1)}(z)\right]
\end{equation}
Since there are two invariant tensors in the su$(2)$ sector ($i=1,2$),
and three in the osp$(2|2)$ sector ($j=1,2,3$), this defines a total
of six
functions $E_{ij}$ upto invididual normalization. In our explicit
evaluation of these embedding blocks, we are able to utilize the
elegant identity in the theory of elliptic integrals which is now
termed Legendre's relation \cite{Legendre:Calcul,Whittaker:Modern}:
\begin{equation}
\label{legrel}
E(z)K(1-z)+E(1-z)K(z)-K(z)K(1-z)=\frac{\pi}{2}
\end{equation}
Straightforward substitution of the su$(2)_0$ and osp$(2|2)_{-2}$
conformal blocks into (\ref{eij}) reveals that the elliptic integrals
always appear in the manner prescribed in Legendre's relation (\ref{legrel}) --- not only are the embedding
blocks single-valued, but they do not contain any elliptic
integrals. Absorbing numerical factors into the coefficients
$c_{ij}$, and defining $E_{ij}=c_{ij}e_{ij}$, the renormalized embedding blocks may be summarized:
\begin{equation}
\label{eij2}
e_{11}=1, \quad e_{12}=e_{13}=\frac{1}{z}, \quad e_{21}=\frac{z}{1-z}, \quad
e_{22}=\frac{1}{1-z},  \quad e_{23}=0
\end{equation}
In subsection \ref{ospfree} we shall present the free field
correlation functions of the  osp$(4|4)_1$ WZNW model, and demonstrate
how one may recover
these results by combining the embedding blocks with the tensorial structure
of osp$(2|2)$ and su$(2)$. 
\subsection{Free Fields and  osp$(4|4)_1$}
\label{ospfree}
The current algebra osp$(4|4)_1$ admits a simple representation in
terms of free fields \cite{Bernard:spincharge}. We introduce the
complex fields $\phi_{\alpha}^{i}$, and their Hermitian conjugates ${\phi_{\alpha}^{i}}^\dagger$, where $i=1,2$ is a species
index, and the index $\alpha=1,2$ denotes boson and fermion
respectively. These eight fields have the following non-trivial
operator product expansions
\begin{equation}
\label{phiopes}
\phi_{\alpha_1}^{i_1}(z){\phi_{\alpha_2}^{i_2}}^\dagger(w)  = 
\frac{\delta^{i_1,i_2}\delta^{\alpha_1,\alpha_2}}{z-w}, \quad {\phi_{\alpha_1}^{i_1}}^\dagger(z){\phi_{\alpha_2}^{i_2}}(w)  = 
(-1)^{\alpha_1}\frac{\delta^{i_1,i_2}\delta^{\alpha_1,\alpha_2}}{z-w}
\end{equation}
and furnish a representation of osp$(4|4)_1$. Let us consider the
holomorphic correlation function
\begin{equation}
\label{fialpha}
{\mathcal F}^{i}_{\alpha}(z_1,\cdots,z_4)=\langle{\phi^{i_1}_{\alpha_1}}(z_1){\phi^{i_2}_{\alpha_2}}^\dagger(z_2){\phi^{i_3}_{\alpha_3}}^\dagger(z_3){\phi^{i_4}_{\alpha_4}}(z_4)\rangle
\end{equation}
where $i$ denotes the ordered sequence $i_1,i_2,i_3,i_4$ and similarly
$\alpha$
denotes $\alpha_1,\alpha_2,\alpha_3,\alpha_4$. Conformal invariance
restricts this correlation function to be of the form
\begin{equation}
\label{confinvfree}
{\mathcal F}^{i}_{\alpha}(z_1,\cdots,z_4)=\left[z_{14}z_{23}\right]^{-2h}{F}^{i}_{\alpha}(z)
\end{equation}
where in this case $h=1/2$. Setting $z_2=0$, $z_3=1$, $z_4=\infty$,
and correspondingly $z_1=z$, one may extract the conformal block
$F^{i}_{\alpha}(z)$:
\begin{equation}
{F}^{i}_{\alpha}(z)=\lim_{w\rightarrow\infty}w\,\langle{\phi^{i_1}_{\alpha_1}}(z){\phi^{i_2}_{\alpha_2}}^\dagger(0){\phi^{i_3}_{\alpha_3}}^\dagger(1){\phi^{i_4}_{\alpha_4}}(w)\rangle
\end{equation}
This correlation function may be evaluated with the aid of Wick's
theorem
\begin{eqnarray}
{F}^{i}_{\alpha}(z)& = &
\lim_{w\rightarrow\infty}w\,\left[\langle{\phi^{i_1}_{\alpha_1}}(z){\phi^{i_2}_{\alpha_2}}^\dagger(0)\rangle\langle{\phi^{i_3}_{\alpha_3}}^\dagger(1){\phi^{i_4}_{\alpha_4}}(w)\rangle
\right. \notag \\
 & &
\hspace{2cm}\left.
+(-1)^{(\alpha_2-1)(\alpha_3-1)}\langle{\phi^{i_1}_{\alpha_1}}(z){\phi^{i_3}_{\alpha_3}}^\dagger(1)\rangle\langle{\phi^{i_2}_{\alpha_2}}^\dagger(0){\phi^{i_4}_{\alpha_4}}(w)\rangle\right]
\end{eqnarray}
where the parity factor $(-1)^{(\alpha_2-1)(\alpha_3-1)}$ contributes
a minus sign if both fields $2$ and $3$ are fermionic . Using the operator product expansions given in (\ref{phiopes}) one obtains
\begin{eqnarray}
\label{fialphafree}
{F}^{i}_{\alpha}(z) & = & \frac{(-1)^{\alpha_3-1}}{z}I_{1,\,{\rm su}}^{i}\,\delta^{\alpha_1,\alpha_2}\delta^{\alpha_3,\alpha_4}+\frac{(-1)^{\alpha_3(\alpha_2-1)}}{z-1}I_{2,\,{\rm su}}^{i}\,\delta^{\alpha_1,\alpha_3}\delta^{\alpha_2,\alpha_4}
\end{eqnarray}
We shall now demonstrate how one may recover this result from the
decomposition
\begin{equation}
\label{embeddecomp}
F^{i}_{\alpha}(z)=\sum_{j=1}^{2}\sum_{k=2}^{3}I_{j,\,{\rm
su}}^{i}I_{k,\,{\rm osp}}^{\alpha^\dagger}c_{jk}e_{jk}(z)
\end{equation}
where $\alpha^\dagger$ denotes the ordered sequence of indices
$\alpha_1,\alpha_2^\dagger,\alpha_3^\dagger,\alpha_4$, and we have
introduced the conjugate indices $1^{\dagger}=4$ and $2^\dagger=3$. We
note in particular that the sum over the osp$(2|2)$ invariant tensors begins at
$k=2$ since the explicit form (\ref{ione}) reveals that
$I_1^{\alpha^\dagger}=0$ for $\alpha_i=1,2$. As we demonstrate in
Appendix \ref{crossing}, crossing symmetry demands that
$c_{22}=-c_{12}$ and $c_{13}=-4\epsilon\gamma c_{12}$. Substituting
these explicit relationships into (\ref{embeddecomp}) along with the
$e_{ij}$ given in equation (\ref{eij2}) one finds
\begin{equation}
\label{fialphareduced}
F^{i}_{\alpha}(z)=c_{12}\left[I_{1,\,{\rm su}}^{i}\left(I_{2,\,{\rm
osp}}^{\alpha^\dagger}-4\epsilon\gamma I_{3,\,{\rm osp}}^{\alpha^\dagger}\right)\frac{1}{z}+I_{2,\,{\rm su}}^{i}I_{2,\,{\rm
osp}}^{\alpha^\dagger}\,\frac{1}{z-1}\right]
\end{equation}
Now, using the explicit form of the osp tensors (\ref{itwo}) and
(\ref{ithree}) one may easily verify that
\begin{align}
\left(I_{2,\,{\rm
osp}}^{\alpha^\dagger}-4\epsilon\gamma I_{3,\,{\rm
osp}}^{\alpha^\dagger}\right)& =
(-1)^{\alpha_3-1}(4\epsilon\gamma)^g\delta_{\alpha_1,\alpha_2}\delta_{\alpha_3,\alpha_4}\\
I_{2,\, {\rm
osp}}^{\alpha^\dagger}  & =  (-1)^{\alpha_3(\alpha_2-1)}(4\epsilon\gamma)^g\delta_{\alpha_1,\alpha_3}\delta_{\alpha_2,\alpha_4}
\end{align}
where $g$ is given by $g=\sum_{i=1}^4\alpha_i/2-2$, one sees
that up to an irrelevant normalization, equation (\ref{fialphareduced}) reduces to
(\ref{fialphafree}). That is to say, we have suceeded in braiding the
su$(2)_0$ and the osp$(2|2)_{-2}$ conformal blocks, in a manner
prescribed by single-valuedness and crossing symmetry, so as to
recover the free field correlation function of osp$(4|4)_1$.
\section{Conclusions}
In this paper we have demonstrated how one may braid the logarithmic conformal
blocks of the su$(2)_{0}$ and osp$(2|2)_{-2}$ WZNW models so as to
recover the free field correlation functions of the osp$(4|4)_1$ model. It is an interesting
open problem to relate these observations to the Hilbert space
non-factorization arguments presented in \cite{Bernard:spincharge}.
\section*{Acknowledgements}The author would like to thank J.-S. Caux,
V. Gurarie, I. I. Kogan, C. Pepin, A. M. Tsvelik, for stimulating
and valuable discussions. The author would also like to thank EPSRC for
financial support.

\appendix
\section{Hypergeometric Series and Elliptic Integrals}
\subsection{Generalities and su$(2)_0$}
We gather here some useful properties of the hypergeometric series and
their connections to the complete elliptic integrals. A good general
reference is \cite{Whittaker:Modern}. More details may
be found in the monographs \cite{Bailey:generalized,Slater:generalized} or in
the various books on special functions
\cite{Prudnikov,Gradshteyn,Bateman,Abromowitz:Tables}. A generalized
hypergeometric series is of the form (see \cite{Gradshteyn} 9.14(1))
\begin{equation}
\label{genhypseries}
\ _pF_q\left[a_1,\ldots,a_p;b_1,\ldots,b_q;z\right]=\sum_{n=0}^\infty \frac{(a_1)\ldots(a_p)_n}{(b_1)_n\ldots(b_q)_n}\frac{z^n}{n!}
\end{equation}
where the so-called Pochammer symbol is defined as
\begin{equation}
(a)_n=a(a+1)\cdots(a+n-1)=\frac{\Gamma(a+n)}{\Gamma(a)}
\end{equation}
When $p=2$ and $q=1$ we obtain the ordinary hypergeometric series. It is well known in the theory of elliptic integrals that (see
\cite{Gradshteyn} 8.113(1) and 8.114(1) and note our different conventions)
\begin{equation}
\label{elliptichyper}
K(z)  =   \frac{\pi}{2}\ _2F_1\left[\tfrac{1}{2},\tfrac{1}{2};1;z\right] \qquad
E(z)  =  \frac{\pi}{2}\ _2F_1\left[-\tfrac{1}{2},\tfrac{1}{2};1;z\right] 
\end{equation}
where $K(z)$ is the complete elliptic integral of the first
kind and $E(z)$ is the complete elliptic integral of the second
kind\footnote{Note that many texts on the theory of elliptic
integrals (including Gradshteyn and Ryzhik \cite{Gradshteyn}) denote the parameter $z$ by
$k^2$ - the so-called modulus. Although this is
purely a mater of convention, we find it preferable to deal with
functions of $z$, rather than $k^2$, or indeed $\sqrt k$.} (see \cite{Gradshteyn} 8.110(1) and the
subsequent discussion)
\begin{equation}
\label{ellipticintegraldefs}
K(z)  =\int_0^1\frac{1}{\sqrt{(1-x^2)(1-zx^2)}}\, dx \qquad E(z)  =\int_0^1\frac{\sqrt{1-zx^2}}{\sqrt{1-x^2}}dx 
\end{equation}
The elliptic integrals have rather simple properties under
differentiation with respect to the parameter $z$ (see
\cite{Gradshteyn} 8.123(2) and 8.132(4) and note our different conventions):
\begin{equation}
\label{simplediff}
\frac{dK(z)}{dz}  =\frac{E(z)-(1-z)K(z)}{2z(1-z)} \qquad \frac{dE(z)}{dz}
 =\frac{E(z)-K(z)}{2z} 
\end{equation}
Given the relations (\ref{elliptichyper}) it is possible to express many other
hypergeometric functions in terms of the elliptic integrals. In
particular, the definiton (\ref{genhypseries}) implies that
\begin{equation}
\ _2F_1^{'}\left[a_1,a_2;b_1;z\right]=\frac{a_1a_2}{b_1}\ _2F_1\left[a_1+1,a_2+1;b_1+1;z\right]
\end{equation}
and thus from (\ref{elliptichyper}) and (\ref{simplediff}) one obtains
\begin{equation}
\label{su0soln1}
\ _2F_1\left[\tfrac{1}{2},\tfrac{3}{2};2;z\right]=-\frac{4}{\pi z}\left[E(z)-K(z)\right]
\end{equation}
Further, using the Gauss recursion formula (see \cite{Gradshteyn} 9.137(3))
\begin{gather}
 (2\beta-\gamma-\beta z+\alpha z)\
_2F_{1}\left[\alpha,\beta;\gamma;z\right]
 +(\gamma-\beta)\ _2F_{1}\left[\alpha,\beta-1;\gamma;z\right]\notag \\
+\beta(z-1)\ _2F_{1}\left[\alpha,\beta+1;\gamma;z\right]=0
\end{gather}
with $\alpha=\beta=1/2$, $\gamma=1$, one finds that
\begin{equation}
\ _2F_1\left[\tfrac{1}{2},\tfrac{3}{2};1;z\right]=(1-z)^{-1}\ _2F_1\left[\tfrac{1}{2},-\tfrac{1}{2};1;z\right]
\end{equation}
Recalling (\ref{elliptichyper}) we may recast this in terms of elliptic integrals
\begin{equation}
\label{su0soln2}
\ _2F_1\left[\tfrac{1}{2},\tfrac{3}{2};1;z\right]=\frac{2E(z)}{\pi(1-z)}
\end{equation}
\subsection{Quadratic Transformations and osp$(2|2)_{-2}$}
\label{quadraticosp}
Amongst the many relations satisfied by the ordinary hypergeometric
functions are the quadratic transformations
\begin{equation}
\ _2F_1[\alpha,\beta;\alpha+\beta+\tfrac{1}{2};4z(1-z)]=
\begin{cases}\ _2F_1[2\alpha,2\beta;\alpha+\beta+\tfrac{1}{2};z] &
\text{(a)} \\ \ _2F_1[2\alpha,2\beta;\alpha+\beta+\tfrac{1}{2};1-z] &
\text{(b)}
\end{cases}
\end{equation}
where form (a) is valid inside the loop of the lemniscate
$|4z(1-z)|=1$ surrounding $z=0$, and form (b) is valid inside the
loop surrounding $z=1$ --- see for example 4.\,(iii) and 4.\,(v) on page 97 of
\cite{Bailey:generalized}. In particular, when applied to the
ancilliary function appearing in (\ref{fom}) with $x=4$ (and $\alpha=-1/4$,
$\beta=3/4$, $\gamma=1$) one obtains 
\begin{equation}
\label{twofom}
F_{0}^{-}=
\begin{cases}
[z(1-z)]^{-1/4}\ _{2}F_{1}\left[-\tfrac{1}{2},\tfrac{3}{2};1;z\right]
& \text{(a)} 
\\
{} [z(1-z)]^{-1/4} \ _{2}F_{1}\left[-\tfrac{1}{2},\tfrac{3}{2};1;1-z\right] 
& \text{(b)}
\end{cases}
\end{equation}
%That there are two independent solutions may also be seen from the
%fact that the ancilliary equation 67 appearing in
%\cite{Maassarani:non} is invariant under the the transformation
%$z\rightarrow1-z$.
Using the Gauss recursion formula (see \cite{Gradshteyn} 9.137(4))
\begin{gather}
\gamma \ _{2}F_{1}\left[\alpha,\beta-1;\gamma;z\right]-\gamma \
_{2}F_{1}\left[\alpha-1,\beta;\gamma;z\right]+(\alpha-\beta)z\ _{2}F_{1}\left[\alpha,\beta;\gamma+1;z\right]=0
\end{gather}
with $\alpha=1/2$, $\beta=3/2$, $\gamma=1$, 
%one finds that
%\begin{equation}
%\ _{2}F_{1}\left[-\tfrac{1}{2},\tfrac{3}{2};1;z\right]=\ _{2}F_{1}\left[\tfra%c{1}{2},\tfrac{1}{2};1;z\right]-z\ _{2}F_{1}\left[\tfrac{1}{2},\tfrac{3}{2};2%;z\right]
%\end{equation}
together with the results  (\ref{elliptichyper}) and (\ref{su0soln1}) one finds that
\begin{equation}
\ _{2}F_{1}\left[-\tfrac{1}{2},\tfrac{3}{2};1;z\right]= \frac{2}{\pi}\left[2E(z)-K(z)\right]
\end{equation}
Substituting this result into (\ref{twofom}) and recalling that the
solutions to the osp$(2|2)_{-2}$ Knizhnik--Zamolodchikov equation (\ref{f3}) are
obtained through the differential relation $F_{3}\propto(1-4D)F_{0}^{-}$, one may utilize
(\ref{simplediff}) and arrive at
the solutions (\ref{ellipticsolutions}) stated in the text.
\section{Invariant Tensors For osp$(2|2)$}
\label{invariant}
The non-vanishing components of
the three invariant tensors for the $[0,1/2]$ representation of
osp$(2|2)$ are equal to the components of the three
vectors \cite{Maassarani:non}
\begin{subequations}
\bea
\label{ione}
I_1 & = & (1144)+(1234)4  \ep\ga+(1324)4  \ep\ga-(1414)+(2143)4  \ep\ga 
\nonumber\\&+&(2233)16 \ep^2\ga^2+
(2323)16 \ep^2\ga^2-(2413)4  \ep\ga+(3142)4  \ep\ga 
+(3232)16 \ep^2\ga^2 \nonumber \\
&+& (3322)16 \ep^2\ga^2 -(3412)4  \ep\ga-(4141)
-(4231)4  \ep\ga-(4321)4  \ep\ga+(4411)\\
& & \nonumber \\
\label{itwo}
I_2&=& (1234)4  \ep\ga-(1243)4  \ep\ga+(1324)4  \ep\ga-(1342)4  
\ep\ga \nonumber\\
&-& (1414)+(1441)-(2134)4  \ep\ga +(2143)4  \ep\ga+(2233)32 \ep^2\ga^2
+(2323)16 \ep^2\ga^2 \nonumber\\
&+& (2332)16 \ep^2\ga^2-(2413)4  \ep\ga
+(2431)4  \ep\ga-(3124)4  \ep\ga+(3142)4  \ep\ga \nonumber\\
&+& (3223)16 \ep^2\ga^2+(3232)16 \ep^2\ga^2
+(3322)32 \ep^2\ga^2-(3412)4  \ep\ga + (3421)4  \ep\ga \nonumber\\
&+&(4114) - (4141)+(4213)4  \ep\ga
-(4231)4  \ep\ga+(4312)4  \ep\ga-(4321)4  \ep\ga\\
&  & \nonumber \\
\label{ithree}
I_3&=&(1234)-(1243)+(1324)-(1342)+(1423)+(1432)-(2134)+(2143)
\nonumber\\ 
&+&(2233)8  \ep\ga +(2314)+(2323)8  \ep\ga+(2332)8  \ep\ga
-(2341)-(2413)+(2431) \nonumber \\
&-& (3124)+(3142)+(3214)+(3223)8  \ep\ga+(3232)8  \ep\ga-(3241)
+(3322)8  \ep\ga \nonumber \\
&-& (3412)+(3421)-(4123)-(4132)+(4213)-(4231)+(4312)-(4321)
\eea
\end{subequations}
For example one finds that $I_1^{3412}=-4\ep\ga$ whilst $I_1^{1441}=0$.
\section{Crossing Symmetry}
\label{crossing}
The correlation function (\ref{fialpha}) has the following transformation under
the permutation of the fields $2$ and $3$ (coordinates and indices) --- crossing symmetry:
\begin{equation}
{\mathcal F}^{i}_{\alpha}(z_1,z_2,z_3,z_4)={\tilde {\mathcal P}}{\mathcal F}^{\tilde{i}}_{\tilde{\alpha}}(z_1,z_3,z_2,z_4)
\end{equation}
where ${\alpha}$ denotes the sequence of indices $\alpha_1,\alpha_2,\alpha_3,\alpha_4$, ${\Tilde\alpha}$ denotes the permuted sequence of
indices $\alpha_1,\alpha_3,\alpha_2,\alpha_4$ (and similarly for $i$),
and ${\Tilde
{\mathcal P}}=(-1)^{(\alpha_2-1)(\alpha_3-1)}$ denotes the parity
of the interchange. Interchanging $z_2$ and $z_3$ induces the
transformation $z\rightarrow 1-z$ (\ref{anhar}) and thus by equation (\ref{confinvfree}),
the crossing symmetry constraint may be also be written
\begin{equation}
F^{i}_{\alpha}(z)=-{\Tilde {\mathcal P}}F^{{\Tilde i}}_{{\Tilde \alpha}}(1-z)
\end{equation}
Returning to our embedding decomposition (\ref{embeddecomp}) and
introducing the following tensors (see equation A.5 of \cite{Maassarani:non})
\begin{equation}
J_{j,\,\rm{su}}^i=I_{j,\,{\rm su}}^{\Tilde i}, \quad J_{h,\,{\rm osp}}^{{\alpha}}={\Tilde {\mathcal P}}
I_{k,\,{\rm osp}}^{\Tilde{{\alpha}}}
\end{equation}
the crossing symmetry constraint takes the form
\begin{equation}
\label{crosssymm2}
\sum_{j=1}^{2}\sum_{k=2}^{3}I_{j,\,{\rm
su}}^{i}I_{k,\,{\rm osp}}^{\alpha^\dagger}c_{jk}e_{jk}(z)=-\sum_{j=1}^{2}\sum_{k=2}^{3}J_{j,\,{\rm
su}}^{{i}}J_{k,\,{\rm osp}}^{\alpha^\dagger}c_{jk}e_{jk}(1-z)
\end{equation}
These new tensors admit the decompositions\footnote{These relations
for osp$(2|2)$ follow from equation A.6 of \cite{Maassarani:non} with
the additional fact that $I_{1,\,{\rm
osp}}^{\alpha_1,\alpha_2^\dagger,\alpha_3^\dagger,\alpha_4}=0$ for
$\alpha_{i}=1,2$ with the identifications  $1^\dagger=4$, $2^\dagger=3$.}
\begin{equation}
J_{1,\,{\rm su}}=I_{2,\,{\rm su}}, \quad J_{2,\,{\rm su}}=I_{1,\,{\rm
su}}, \quad J_{1,\,{\rm osp}}=I_{2,\,{\rm osp}}-4\epsilon\gamma
I_{3,\,{\rm osp}}, \quad J_{3,\,{\rm osp}}=-I_{3,\,{\rm osp}}
\end{equation}
Substituting these decompositions into equation (\ref{crosssymm2}) and
equating the coefficients of $I_{j,\,{\rm su}}I_{k,\,{\rm osp}}$ on both sides, one
finds the following identities which must be satisfied by the
combinations $c_{ij}e_{ij}(z)$ if crossing symmetry is to be
satisfied:
\begin{subequations}
\begin{eqnarray}
c_{12}e_{12}(z) & = & -c_{22}e_{22}(1-z) \label{c12}\\
c_{13}e_{13}(z) & = & 4\epsilon\gamma
c_{22}e_{22}(1-z)+c_{23}e_{23}(1-z)\label{c13}\\
c_{22}e_{22}(z) & = & -c_{12}e_{12}(1-z)\label{c22}\\
c_{23}e_{23}(z) & = & 4\epsilon\gamma c_{12}e_{12}(1-z)+c_{13}e_{13}(1-z)\label{c23}
\end{eqnarray}
\end{subequations}
Inserting the explicit expressions for $e_{ij}$ (\ref{eij2}) one finds
that (\ref{c12}) implies $c_{22}=-c_{12}$, and 
(\ref{c13}) implies $c_{13}=4\epsilon\gamma c_{22}$. Equations
(\ref{c22}) and (\ref{c23}) simply reproduce these
identifications.
%\bibliographystyle{/home/wytham/bhaseen/Refs/h-physrev}
%\bibliography{/home/wytham/bhaseen/Refs/refs}

\end{document}